\documentclass[12pt]{iopart}

\usepackage{iopams}

\begin{document}
\hspace{5.2in} \mbox{IPMU13-0072} 

\title{On the cosmology of massive gravity}

\author{Antonio De Felice}
\address{ThEP's CRL, NEP, The Institute for Fundamental Study, Naresuan University,
Phitsanulok 65000, Thailand}
\address{Thailand Center of Excellence in Physics, Ministry of Education, Bangkok 10400, Thailand}
\ead{antoniod@nu.ac.th}

\author{A. Emir G\"umr\"uk\c{c}\"uo\u{g}lu}
\address{Kavli Institute for the Physics and Mathematics of the Universe (WPI), \\Todai Institutes for Advanced Study, University of Tokyo,\\ 5-1-5 Kashiwanoha, Kashiwa, Chiba 277-8583, Japan}
\ead{emir.gumrukcuoglu@ipmu.jp}

\author{Chunshan Lin}
\address{Kavli Institute for the Physics and Mathematics of the Universe (WPI), \\Todai Institutes for Advanced Study, University of Tokyo,\\ 5-1-5 Kashiwanoha, Kashiwa, Chiba 277-8583, Japan}
\ead{chunshan.lin@ipmu.jp}

\author{Shinji Mukohyama}
\address{Kavli Institute for the Physics and Mathematics of the Universe (WPI), \\Todai Institutes for Advanced Study, University of Tokyo,\\ 5-1-5 Kashiwanoha, Kashiwa, Chiba 277-8583, Japan}
\ead{shinji.mukohyama@ipmu.jp}

\begin{abstract}
 We present a review of cosmological solutions in non-linear massive
 gravity, focusing on the stability of perturbations. Although
 homogeneous and isotropic solutions have been found, these are now
 known to suffer from either the Higuchi ghost or a new non-linear ghost
 instability. We discuss two approaches to alleviate this issue. By
 relaxing the symmetry of the background by e.g.\ 
 breaking isotropy in the hidden sector, it is possible to
 accommodate a stable cosmological solution. Alternatively, extending
 the theory to allow for new dynamical degrees of freedom can also
 remove the conditions 
which lead to the instability. As examples for
 this case, we study the stability of self-accelerating solutions in
 the quasi-dilatonic extension and generic cosmological solutions in
 the varying mass extension. While the quasi-dilaton case turns out to
 be unstable, the varying mass case allows stable regimes of
 parameters. Viable self-accelerating solutions in the varying mass
 theory yet remain to be found.
\end{abstract}

\maketitle

\section{Introduction}

The search for a finite-range gravity has been a long-standing problem, well motivated by both theoretical and
observational considerations. On the theory side, the existence of a theoretically consistent extension of general relativity (GR) by a mass term has been a basic question of classical field theory. After Fierz and
Pauli's pioneering attempt in 1939~\cite{Fierz:1939ix}, this issue has 
attracted a great deal of interest. On the observation side,
continuing experimental probes of gravity have revealed new unexpected
phenomena at large scales; one of the most profound discovery is the
cosmic acceleration, which was found in 1998 \cite{SNe}. The extremely tiny
energy-scale associated with the cosmic acceleration hints that gravity
might need to be modified in the infrared (IR). The massive gravity is
one of the most interesting attempts in this direction.

However, theoretical and observational consistency of massive gravity theories has been a
challenging issue for several decades. Fierz and Pauli's 
model~\cite{Fierz:1939ix}, which extends GR by a linear mass term, 
suffers from the van Dam-Veltman-Zakharov
discontinuity~\cite{vanDam:1970vg, Zakharov:1970cc}; relativistic and non-relativistic
matter couple to gravity with different relative
strengths, no matter how small the graviton mass is. Although this
problem can be alleviated by non-linear effects, as suggested by
Vainshtein~\cite{Vainshtein:1972sx}, the same non-linearities lead to a
ghost instability. Indeed, at the non-linear level, 
the theory loses not
only the momentum constraint but also the Hamiltonian constraint and, as
a result, the non-linear theory includes up to $6$ degrees of freedom in
the gravity sector. While $5$ of them properly represent the degrees of
freedom of a massive spin-$2$ field in a Poincar{\'e} 
 invariant background,
the sixth one is the so called Boulware-Deser (BD)
ghost~\cite{Boulware:1973my}.

Adopting the effective field theory approach in the decoupling limit (i.e. $m_g\to 0$, $M_p\to \infty$, $\Lambda\to {\rm fixed}$, where $\Lambda$ is the cut-off of the theory), it
was found that the BD ghost is related to the longitudinal mode of the
Goldstone bosons associated with the broken general
covariance~\cite{ArkaniHamed:2002sp}.Construction of a theory free from 
the BD
ghost was only recently achieved by de Rham, Gabadadze and
Tolley (dRGT)~\cite{deRham:2010ik,deRham:2010kj}. It was shown that the 
Hamiltonian constraint and the associated secondary constraint are
restored in this theory, eliminating the BD ghost mode as a result~\cite{Hassan:2011hr,deRham:2011rn,deRham:2011qq,Hassan:2011tf,Mirbabayi:2011aa,Golovnev:2011aa,Kluson:2012wf}. \footnote{See Ref.\cite{bimetric} for the proof of the absence of the BD degree in the bi-metric and multi-metric extensions.}

However, in order for the theory to be theoretically consistent and
observationally viable, the absence of the
BD ghost is not
sufficient. At the very least, a stable cosmological solution is needed.

The purpose of this article is to review the construction and the
stability of cosmological solutions in the context of non-linear massive
gravity. We start with describing the action of dRGT theory in
Sec.~\ref{sec:action}.  In Sec.~\ref{sec:FLRW} we construct
homogeneous and isotropic cosmological solutions that exhibit
self-acceleration. We then argue in Sec.~\ref{sec:ghost} that all
homogeneous and isotropic solutions in the dRGT theory are unstable and
thus cannot describe the universe as we know it. In
Sec.~\ref{sec:alternatives} we propose three alternative cosmological
scenarios to avoid instabilities. One of them is based on the
observation that breaking isotropy in the hidden sector (fiducial
metric) still allows isotropic evolution of the visible sector (physical
metric) and thus the standard thermal history of the universe. The other
two proposals maintain isotropy in both visible and hidden sectors but
are based on extended theories of massive gravity with extra degrees of
freedom, such as the quasi-dilaton theory~\cite{D'Amico:2012zv} and the
varying mass theory~\cite{D'Amico:2011jj,Huang:2012pe}. \footnote{We note that another possible extension, not considered here, is the bi-metric theory \cite{bimetric}, where the fiducial metric is promoted to a second, dynamical metric. The cosmology \cite{vonStrauss:2011mq, Comelli:2011zm} allows self-acceleration \cite{Akrami:2012vf}. The cosmological perturbations was studied in \cite{Comelli:2012db}.}

\section{Action}
\label{sec:action}

We start by describing the action of dRGT massive gravity
theory~\cite{deRham:2010kj}. In order to have a manifestly
diffeomorphism invariant description of the massive gravity, the action
is built out of four St\"uckelberg scalar fields,
$\phi^a(x),~a=0,1,2,3$. These four scalars enter the gravity action
through a ``fiducial metric'' defined as 
\begin{eqnarray}
f_{\mu\nu}\equiv \bar{f}_{ab}(\phi^c)
 \partial_{\mu}\phi^a\partial_{\nu}\phi^b~, \label{eqn:fmunu}
\end{eqnarray}
where the ``reference metric'' $\bar{f}_{ab}(\phi^c)$ is a metric in the field space. The 
action for the gravity sector is a functional of the physical metric
$g_{\mu\nu}$ and the fiducial metric $f_{\mu\nu}$. A necessary condition 
for the theory to be free from BD ghost is that the action in the
decoupling limit to vanish up to boundary terms when restricted to the
longitudinal part of the St\"uckelberg fields. With this requirement,
the most general mass term without derivatives of $g_{\mu\nu}$ and
$f_{\mu\nu}$ is constructed as 
\begin{eqnarray}
 S_{mass}[g_{\mu\nu},f_{\mu\nu}] & = & 
  M_{Pl}^2m_g^2\int d^4x\sqrt{-g}\,
  ( {\cal L}_2+\alpha_3{\cal L}_3+\alpha_4{\cal L}_4), 
  \label{eqn:Imass}
\end{eqnarray}
with
\begin{eqnarray}
 {\cal L}_2 & = & \frac{1}{2}
  \left(\left[{\cal K}\right]^2-\left[{\cal K}^2\right]\right)\,, \quad
 {\cal L}_3 = \frac{1}{6}
  \left(\left[{\cal K}\right]^3-3\left[{\cal K}\right]\left[{\cal K}^2\right]+2\left[{\cal K}^3\right]\right), 
  \nonumber\\
 {\cal L}_4 & = & \frac{1}{24}
  \left(\left[{\cal K}\right]^4-6\left[{\cal K}\right]^2\left[{\cal K}^2\right]+3\left[{\cal K}^2\right]^2
   +8\left[{\cal K}\right]\left[{\cal K}^3\right]-6\left[{\cal K}^4\right]\right)\,,
\label{lag234}
\end{eqnarray}
where a square bracket denotes trace operation and 
\begin{equation}
{\cal K}^\mu _\nu = \delta^\mu _\nu 
 - \left(\sqrt{g^{-1}f}\right)^{\mu}_{\ \nu}\,.
\label{Kdef}
\end{equation} 
It was shown that the theory is free from BD ghost at the fully
non-linear level even away from the decoupling
limit~\cite{Hassan:2011hr,deRham:2011rn,deRham:2011qq,Hassan:2011tf}.

\section{FLRW cosmological solution}
\label{sec:FLRW}

With the non-linear massive gravity theory free from BD ghost in hand,
it is important to study its cosmological implications. In this section
we thus construct Friedmann-Lema\^\i tre-Robertson-Walker (FLRW)
solutions.

\subsection{Open FLRW solution with Minkowski reference metric}
\label{subsec:openFLRW}

The original dRGT theory respects the Poincar{\'e} symmetry in the field
space and thus the reference metric is Minkowski, i.e.\
$\bar{f}_{ab}=\eta_{ab}={\rm diag}(-1,1,1,1)$. In this subsection we
thus review the FLRW solution with the Minkowski reference
metric~\cite{Gumrukcuoglu:2011ew}. This is the first non-trivial FLRW
solution in the context of dRGT massive gravity.

In order to find FLRW cosmological solutions we should adopt an ansatz
in which both $g_{\mu\nu}$ and $f_{\mu\nu}$ respect the FLRW
symmetry. Since the tensor $f_{\mu\nu}$ is the pullback of the Minkowski 
metric in the field space to the physical spacetime, such an ansatz
would require a flat, closed, or open FLRW coordinate system for the 
Minkowski line element. The Minkowski line element does not admit
a closed chart, but it allows an open chart. Thus, while there is no
closed FLRW solution, we may hope to find open FLRW solutions. A flat
FLRW solution, if it
exists, is on the boundary between closed and open
solutions but it was shown in \cite{D'Amico:2011jj} 
that such a solution does not exist. For these reasons, in the following 
we shall seek open FLRW solutions.

Motivated by the coordinate transformation from Minkowski coordinates to
Milne coordinates, we take the following ansatz for the four
St\"uckelberg scalars. 
\begin{equation}
\phi^0 = f(t)\sqrt{1+|K|\delta_{ij}x^ix^j}~,\quad
\phi^i =\sqrt{|K|}f(t)x^i~,
\end{equation}
with $K<0$. This leads to the open FLRW form for the Minkowski fiducial
metric, 
\begin{eqnarray}
f_{\mu\nu}\equiv\eta_{ab}\partial_{\mu}\phi^a\partial_{\nu}\phi^b=-\left(\dot{f}(t)\right)^2\delta^0_{\mu}\delta^0_{\nu}+|K|f(t)^2\Omega_{ij}\delta^i_{\mu}\delta^j_{\nu}~.
\end{eqnarray}
As for the physical metric, we adopt the general open ($K<0$) FLRW ansatz as 
\begin{eqnarray}
g_{\mu\nu}dx^{\mu}dx^{\nu}&=&-N(t)^2dt^2+\Omega_{ij}dx^idx^j~,\nonumber\\
\Omega_{ij}dx^idx^j&=&dx^2+dy^2+dz^2+\frac{K(xdx+ydy+zdz)^2}{1-K(x^2+y^2+z^2)}~. \label{eqn:FLRW-physicalmetric}
\end{eqnarray}
Here, $x^0=t,~x^1=x,~x^2=y,~x^3=z;~\mu,\nu=0,...,3$ and $i,j=1,2,3$. 
Without loss of generality, we assume that 
$\dot{f}>0~, f>0,~ a>0,~ N>0$. By substituting the above ansatz into the
Einstein-Hilbert action plus the mass term (\ref{eqn:Imass}), the
gravity action up to a boundary term can be written as 
 \begin{equation}
 S_g = \int d^4x\sqrt{\Omega}  \left[-3|K|Na-\frac{3\dot{a}^2a}{N}
      +m_g^2 \left( L_2+\alpha_3L_3+\alpha_4L_4\right) \right]\,,
\label{actIg}
\end{equation}
where 
\begin{eqnarray}
 L_2 & = & 3a(a-\sqrt{|K|}f)(2Na-\dot{f}a-N\sqrt{|K|}f)\,, \nonumber\\
 L_3 & = & (a-\sqrt{|K|}f)^2(4Na-3\dot{f}a-N\sqrt{|K|}f)\,, \nonumber\\
 L_4 & = & (a-\sqrt{|K|}f)^3(N-\dot{f})\,.
\end{eqnarray}
In addition to the gravity action, we also consider a general matter
content so that the total action is $S_{\rm tot}=S_g+S_{\rm matter}$.

Note that since the above ansatz fully respects the FLRW symmetry, the 
$(0i)$ components of the equations of motion for $g_{\mu\nu}$ are
trivially satisfied, thus the variation of the action with respect to
$N(t)$ and $a(t)$ should correctly give all the non-zero components of
the Einstein equation. On the other hand, because of the
identity~\cite{Hassan:2011vm} 
\begin{equation}
 \nabla^{\mu}\left(\frac{2}{\sqrt{-g}}
	      \frac{\delta S}{\delta g^{\mu\nu}}\right)
 = \frac{1}{\sqrt{-g}}
 \frac{\delta S_g}{\delta\phi^a}\partial_{\nu}\phi^a, 
\label{stuckid}
\end{equation}
the number of independent equations of motion for the 
St\"uckelberg scalars is one.

Now let us take the variation of the action with respect to $f(t)$,
which contains all non-trivial information about the dynamics of the
St\"uckelberg scalars. It leads to 
\begin{equation}
 (\dot{a}-\sqrt{|K|}N)
  \left[(3-2X) + \alpha_3(3-X)(1-X) + \alpha_4(1-X)^2 \right]
  = 0,
 \label{eqn:wrtf}
\end{equation}
where $ X\equiv\sqrt{|K|}f/a$. This equation has three solutions. The
first one is $\dot{a}=\sqrt{|K|}N$ and corresponds to an empty open
universe, i.e.\ the open FLRW chart of Minkowski spacetime. Thus this
solution is not of our interest. The remaining two solutions are 
\begin{eqnarray}
f = \frac{a}{\sqrt{|K|}}X_{\pm}, \quad
 X_{\pm} \equiv \frac{1+2\alpha_3+\alpha_4
 \pm\sqrt{1+\alpha_3+\alpha_3^2-\alpha_4}}
 {\alpha_3+\alpha_4}\,.
\label{eqn:f-sol}
\end{eqnarray}
Note that these two solutions are singular in the limit $K\to 0$. This 
is consistent with the result in \cite{D'Amico:2011jj},
i.e.\ the non-existence of flat FLRW cosmologies. On the other hand, with
$K< 0$, by taking the variation of the action with respect to $N(t)$ and
using (\ref{eqn:f-sol}), we obtain the following modified Friedmann
equation. 
\begin{equation}
 3H^2+\frac{3K}{a^2} 
  = \rho_m +  \Lambda_{\pm}, \quad 
  H\equiv \frac{\dot{a}}{Na}, 
  \label{eqn:N-eq-pre}
\end{equation}
where $\rho_m$ is the energy density of the matter sector and 
\begin{eqnarray}
 \Lambda_{\pm} &\equiv &
  -\frac{m_g^2}
  {(\alpha_3+\alpha_4)^2}
  \left[1+\alpha_3\pm\sqrt{1+\alpha_3+\alpha_3^2-\alpha_4}\right]
  \nonumber\\
 & & \qquad\qquad\times
  \left[1+\alpha_3^2-2\alpha_4\pm(1+\alpha_3)
   \sqrt{1+\alpha_3+\alpha_3^2-\alpha_4}\right].
  \label{eqn:Lambdapm}
\end{eqnarray}
In this way the graviton mass manifests as the effective cosmological
constant $\Lambda_{\pm}$. When $\Lambda_{\pm}>0$, the system exhibits
self-acceleration. By taking the variation of the action with respect to
$a(t)$ we obtain a dynamical equation, which is consistent with the
above modified Friedmann equation and the standard conservation equation
for matter.

\subsection{Flat/closed/open FLRW solutions with general reference metric}
\label{subsec:generalFRW}

In appendix of Ref.~\cite{Gumrukcuoglu:2011zh}, the open FLRW
solution was generalized to flat/closed/open FLRW solutions by
considering a fiducial metric of the general FLRW type. (General reference
metrics were first considered in \cite{Hassan:2011vm} and the
absence of BD ghost in this general setup was proven by
\cite{Hassan:2011tf}. See also \cite{bimetric} for the absence of ghost in the bi-metric theory.) In this subsection we describe the general
solutions.

The most general fiducial metric consistent with flat ($K=0$), closed
($K>0$) or open ($K<0$) FLRW symmetries is 
\begin{eqnarray}\label{eqn:generalfid} 
f_{\mu\nu}=-n^2(\varphi^0)\partial_{\mu}\varphi^0\partial_{\nu}\varphi^0+\alpha^2(\varphi^0)\Omega_{ij}(\varphi^k)\partial_{\mu}\varphi^i\partial_{\nu}\varphi^j~,
\end{eqnarray} 
where $n$ and $\alpha$ are general functions of $\varphi^0$, and
$\Omega_{ij}(\varphi^k)$ is defined as in
(\ref{eqn:FLRW-physicalmetric}) with $(x,y,z)$ replaced by
$(\varphi^1,\varphi^2,\varphi^3)$, and the curvature constant $K$ is now
either zero, positive or negative. Here, we have used the notation
$\varphi^a$ instead of $\phi^a$ to make it clear that this form of the
fiducial metric may be achieved from the original form (\ref{eqn:fmunu}) 
by non-trivial change of variables (as we have explicitly seen in the
previous subsection). As for the physical metric, we adopt the ansatz
(\ref{eqn:FLRW-physicalmetric}) with an arbitrary sign for $K$.

Similarly to the case in the previous subsection, the equation of motion
for the St\"uckelberg fields allows three branches of solutions. In the
general setup at hand, the first branch is characterized by 
$aH=\alpha H_f$, where $H\equiv\dot{a}/(Na)$ and 
$H_f\equiv\dot{\alpha}/(n\alpha)$ are Hubble expansion rates of the
physical and fiducial metrics, respectively. Unfortunately, this branch
would not allow nontrivial cosmologies since it does not evade the Higuchi
bound~\cite{Higuchi:1986py} and thus linear perturbations around the
corresponding solution
\cite{Hassan:2011vm,Fasiello:2012rw,Langlois:2012hk} include a ghost degree in
the cosmological history. Therefore, we shall not consider this branch and
restrict our attention to the
other branches.

The two remaining branches are characterized by $\alpha=X_{\pm} a$, where
$X_{\pm}$ is the same as in (\ref{eqn:f-sol}). For these two branches,
the metric equation of motion is exactly the same as
(\ref{eqn:N-eq-pre}) with (\ref{eqn:Lambdapm}). Surprisingly enough, the
modified Friedmann equation (including the value of the effective
cosmological constant induced by the graviton mass term) does not depend
on the properties of the fiducial metric at all. When $\Lambda_{\pm}>0$,
the system exhibits self-acceleration.

\section{New non-linear instability of FLRW solutions}
\label{sec:ghost}

In the previous section we have constructed flat, closed and open FLRW
solutions in non-linear massive gravity with a general FLRW fiducial
metric. The construction allows three branches of solutions. However,
the first branch characterized by $aH=\alpha H_f$ suffers from the
Higuchi ghost at the level of linear perturbations and thus does not
allow a non-trivial cosmological history. In this section we thus
consider the other two branches of solutions characterized by 
$\alpha=X_{\pm} a$. These solutions evade the Higuchi ghost, but
unfortunately we shall see that a new type of ghost instability shows up
at non-linear level~\cite{DeFelice:2012mx}. Based on this result, we
shall argue that all homogeneous and isotropic FLRW solutions in the
dRGT theory are unstable. We shall then propose alternative cosmological 
scenarios in the next section.

\subsection{Linear perturbation}
\label{subsec:linear}

In this subsection, following \cite{Gumrukcuoglu:2011zh}, we shall 
investigate linear perturbations around a general flat/closed/open FLRW
solution (characterized by $\alpha=X_{\pm} a$) with a general FLRW
fiducial metric and an arbitrary
matter content. We shall see that time
kinetic terms for $3$ among the $5$ graviton degrees of freedom always
vanish at the level of the quadratic action, signaling for necessity of
non-linear analysis.

We first define perturbations of four St\"uckelberg scalars through the
exponential mapping, truncating at the second order, as 
\begin{eqnarray}
\varphi^a=x^a+\pi^a+\frac{1}{2}\pi^b\partial_b\pi^a+\mathcal{O}(\pi^3)~.
\end{eqnarray}
We then perturb the physical metric as
\begin{equation}
 g_{00} = -N^2(t)\left[1+2\phi\right], \quad
 g_{0i} = N(t)a(t)\beta_i, \quad
 g_{ij} = a^2(t)\left[\Omega_{ij}+h_{ij}\right]. 
\end{equation} 
We suppose that $\pi^a,\phi,\beta_i,h_{ij}=O(\epsilon)$. The following 
gauge-invariant variables can be constructed out of St\"uckelberg and
metric perturbations. 
\begin{eqnarray}
 \phi^{\pi} & \equiv & \phi - \frac{1}{N}\partial_t(N\pi^0), \quad
 \beta^{\pi}_i \equiv 
  \beta_i + \frac{N}{a}D_i\pi^0 - \frac{a}{N}\dot{\pi}_i, \nonumber\\
 h^{\pi}_{ij} & \equiv & h_{ij} 
  - D_i\pi_j - D_j\pi_i - 2NH\pi^0\Omega_{ij}, 
  \label{eqn:phipi-betapi-hpi}
\end{eqnarray}
where $D_i$ is the spatial covariant derivative compatible with
$\Omega_{ij}$.

In Sec.~\ref{sec:FLRW} we have seen that the mass term acts as an 
effective cosmological constant at the background level. Hence, we
define 
\begin{equation}
 \tilde{S}_{mass}[g_{\mu\nu},f_{\mu\nu}] \equiv 
  S_{mass}[g_{\mu\nu},f_{\mu\nu}]
  +M_{Pl}^2\int d^4x\sqrt{-g}\, \Lambda_{\pm},
  \label{eqn:def-Itildemass}
\end{equation}
where $\Lambda_{\pm}$ is specified in (\ref{eqn:Lambdapm}), and expand
$\tilde{S}_{mass}$ instead of $S_{mass}$. This greatly simplifies the
perturbative expansion. As shown in \cite{Gumrukcuoglu:2011zh}, upon
using the background equation of motion for the St\"uckelberg fields but
without using the background equation of motion for the physical metric,
the quadratic part of $\tilde{S}_{mass}$ is simplified as
\begin{eqnarray} 
 \tilde{S}_{mass}^{(2)} & = & 
  \frac{M_{Pl}^2}{8}\int d^4x Na^3\sqrt{\Omega}\, M_{GW}^2
  \left[ (h^{\pi})^2-h_{\pi}^{ij}h^{\pi}_{ij}\right],
  \label{eqn:Imass2}
\end{eqnarray}
where 
\begin{equation}
 M_{GW}^2 \equiv \pm (r-1)m_g^2\,
  X_{\pm}^2\sqrt{1+\alpha_3+\alpha_3^2-\alpha_4},\quad
  r \equiv \frac{na}{N\alpha} = \frac{1}{X_{\pm}}\frac{H}{H_f}, 
\end{equation}
$X_{\pm}$ is given by (\ref{eqn:f-sol}), 
$h^{\pi}\equiv\Omega^{ij}h^{\pi}_{ij}$,
$h_{\pi}^{ij}\equiv\Omega^{ik}\Omega^{jl}h^{\pi}_{kl}$, and
$\Omega^{ij}$ is the inverse of $\Omega_{ij}$. This is manifestly
gauge-invariant.

What is important here is that the gauge-invariance of
$\tilde{S}_{mass}^{(2)}$ was shown without using the background equation
of motion for the physical metric. This means that
$\tilde{S}_{mass}^{(2)}$ is gauge-invariant for any matter content (as
far as the matter action does not depend on the St\"uckelberg fields so
that the St\"uckelberg equation of motion is derived solely from the
graviton mass term) and that the remaining part 
$\tilde{S}_{\rm GR}^{(2)}\equiv S_{\rm tot}^{(2)}-\tilde{S}_{mass}^{(2)}$ 
of the total (gravity plus matter) quadratic action $S_{\rm tot}^{(2)}$
is also gauge-invariant by itself. Hence, the remaining part
$\tilde{S}_{\rm GR}^{(2)}$ never depends on the St\"uckelberg
perturbations for any gauge choice. Another important point is that 
$\tilde{S}_{mass}^{(2)}$ shown in (\ref{eqn:Imass2}) does not depend on
$\phi^{\pi}$ and $\beta_i^{\pi}$, and hence does not include time
derivatives of St\"uckelberg perturbations. Therefore, for any matter
content, the dependence of the total quadratic action on the
St\"uckelberg perturbations is completely given by (\ref{eqn:Imass2})
and time derivatives of St\"uckelberg perturbations do not enter
the quadratic action at all. This completes the proof of the statement
that time kinetic terms for $3$ among $5$ gravity degrees of freedom
always vanish at the level of the quadratic action. This proof holds for
any matter content~\cite{Gumrukcuoglu:2011zh}.

The absence of quadratic kinetic terms for $3$ gravity degrees of
freedom shown in this subsection implies that the self-accelerating FLRW
solutions evade the Higuchi bound \cite{Higuchi:1986py} and thus are
free from ghost at the linearized level even when the expansion rate is
significantly higher than the graviton mass. At the same time, this signals for the necessity of a non-linear analysis. In contrast, the first branch solution mentioned in Sec.\ref{subsec:generalFRW}, which gives $a\,H=\alpha\,H_f$, contains five propagating degrees of freedom. In this case however, one of these degrees turns out to be the Higuchi ghost \cite{Fasiello:2012rw}.

\subsection{Non-linear perturbation}
\label{subsec:nonilnear}

In order to understand the physical content of the FLRW background in dRGT massive gravity, we need to investigate the reason why the kinetic term of one of the scalar modes and two of the vector ones have a vanishing kinetic term. We will show that this feature does not hold in general for the theory, but is rather a consequence of the symmetries of the FLRW background.

Therefore it is convenient to study another background which has less symmetries than FLRW, but does lead FLRW in some limit. Probably, the simplest implementation of such a background is the axisymmetric Bianchi Type-I class of metrics, which can be written as
\begin{equation}
ds^2 = -N^2(t) dt^2+a^2(t)[e^{4\sigma(t)}dx^2+e^{-2\sigma(t)}\delta_{ij} dy^idy^j]\,,
\end{equation}
where here $i,j\in\{2,3\}$, and $y^2=y,y^3=z$. It is evident that this
manifold is not isotropic, however as $\sigma$ approaches $0$, the
spacetime reduces to flat FLRW. Since we want to study the reason why there
are missing kinetic terms for the perturbed FLRW fields, we will not
consider the above metric as physical. Rather, for the time being,
we will merely use it as a tool to study non-linear perturbations on
FLRW. In other words, linear perturbations on an anisotropic manifold
will give information equivalent to non-linear perturbation theory
on FLRW. This provides a consistent truncation of the non-linear perturbations, allowing us to analyze them in a simple way. The goal of this section is to show how the missing kinetic
terms will depend on $\sigma$, namely we will study their sign and the
dispersion relations of the associated perturbation fields. 

According to the properties of the perturbation fields under a rotation in the $y{-}z$ plane, we can decompose such fields into scalar (a.k.a.\ even) and vector (a.k.a.\ odd) modes. In particular we can write the metric for the even modes as
\begin{eqnarray}
ds_{\rm even}^2 &=& -N^2(1+2\Phi)dt^2 + 2aNdt[e^{2\sigma}\partial_x \chi dx+ e^{-\sigma}\partial_i B dy^i]
\nonumber\\
&&{}+ a^2e^{4\sigma}(1+\psi)dx^2 
+2a^2e^\sigma\partial_x\partial_i\beta dxdy^i \nonumber\\
&&{}+ a^2e^{-2\sigma}[\delta_{ij}(1+\tau) +\partial_i\partial_j E]dy^idy^j\,,
\label{eq:evenmetric}
\end{eqnarray}
whereas the metric for the odd modes reads
\begin{eqnarray}
ds_{\rm odd}^2 &=&-N^2dt^2 +2ae^{-\sigma}Nv_idtdy^i+2a^2e^\sigma\partial_x\lambda_i dxdy^i\nonumber\\
&&{}+a^2e^{4\sigma} dx^2+a^2e^{-2\sigma}(\delta_{ij}+\partial_{(i}h_{j)})dy^idy^j\,.
\label{eq:oddmetric}
\end{eqnarray}
where the transverse condition holds, i.e.\  $\partial^i v_i=0=\partial^i \lambda_i=0=\partial^i h_i$. 
At the same time, we also need to specify the form of the fiducial metric $f_{\mu\nu}$. Since, we want to study the FLRW limit, we need the fiducial metric to possess already the FLRW symmetries from the beginning. The anisotropic physical metric corresponds to the simplest deviation from an overall FLRW symmetry for the whole system. Therefore we assume
\begin{equation}
f_{\mu\nu} = -n(\phi^0)^2 \partial_\mu \phi^0 \partial_\nu \phi^0+ \alpha(\phi^0)^2 (\partial_\mu\phi^1\partial_\nu \phi^1+\delta_{ij}\partial_\mu\phi^i\partial_\nu \phi^j),
\label{eq:fmnB}
\end{equation}
so that we need to give the perturbations of the St\"uckelberg fields according to even/odd mode decomposition.
In fact, the even-type perturbations of St\"uckelberg fields read
\begin{equation}
\phi^0 = t + \pi^0\,,\;
\phi^1 = x + \partial_x\pi^1\,,\;
\phi^i = y^i + \partial^i \pi\,.
\end{equation}
For the odd modes sector, we consider instead 
\begin{equation}
\phi^0=t\,,\,\phi^1=x\,,\,\phi^i=y^i + \pi^i\,,
\label{eq:oddstuckB}
\end{equation}
where  $\partial_i \pi^i=0$. 

It is possible to define gauge-invariant fields for even perturbations as follows 
\begin{eqnarray}
\hat{\Phi} &=& \Phi-
 \frac{1}{2\,N}\,\partial_t\left(\frac{\tau}{H-\Sigma}\right)\,,\quad
\hat{B}= B + \frac{e^\sigma}{2\,a\,(H-\Sigma)}\,\tau - \frac{a\,e^{-\sigma}}{2\,N}\,\dot{E}\,,\nonumber\\
\hat{\chi} &=& \chi + \frac{\tau\, e^{-2\sigma}}{2a(H-\Sigma)}- \frac{ae^{2\sigma}}{N}\partial_t\!\left[ e^{-3\sigma}\!\left(\beta - \frac{e^{-3\sigma}}{2}E\right)\right],\nonumber\\
\hat{\psi} &=& \psi -\frac{H+2\,\Sigma}{H-\Sigma}\,\tau - e^{-3\,\sigma}\,\partial_x^2\left(2\,\beta - e^{-3\,\sigma}\,E\right),\quad
\hat{E}_\pi = \pi -\frac{1}{2}\,E\,, \nonumber\\
 \hat{\tau}_\pi & = & \pi^0 - \frac{\tau}{2\,N\,(H-\Sigma)}\,,\quad
\hat{\beta}_\pi = \pi^1 - e^{-3\,\sigma}\left(\beta - \frac{e^{-3\,\sigma}}{2}\,E\right),
\label{eq:gi-gr-even}
\end{eqnarray}
where we have defined $\Sigma \equiv \dot{\sigma}/{N}$.

Then we can proceed to integrate out all the present auxiliary
fields. In general, we can integrate out three modes, that is
$\hat\Phi,\hat B$ and $\hat\chi$. However, in the dRGT theory, it is
possible to show that also the field $\hat\tau_\pi$ can be integrated
out. Therefore there are only three independent fields describing the
even-mode perturbations, so we need to study the kinetic matrix
of the three remaining fields, $\hat\psi$, $\hat\beta_\pi$, and 
$\hat E_\pi$. As $\sigma\to0$, the eigenvalues of the $3\times3$ kinetic 
matrix reduce to 
\begin{equation}
\kappa_1 \simeq \frac{p_T^4}{8\,p^4}\,,\;
\kappa_2 \simeq - \frac{2a^4M_{\rm GW}^2p_L^2}{1-r^2}\sigma\,,\;
\kappa_3 \simeq -\frac{p_T^2}{2\,p_L^2}\,\kappa_2\,,
\end{equation}
where we have introduced 
$r\equiv an/(\alpha N)$,
$M_{\rm GW}^2 \equiv m_g^2(1-r)X^2[(1+2\alpha_3+\alpha_4)-(\alpha_3+\alpha_4)X]$, $X \equiv{\alpha}/{a}$,
$p_L \equiv {k_L}/({ae^{2\sigma}})\simeq k_L/a$, 
$p_T \equiv {k_T}/(ae^{-\sigma})\simeq k_T/a$, 
$k_T^2\equiv k_y^2+k_z^2$, and 
$p^2 \equiv p_L^2 + p_T^2$. The first and most important consideration is that $\kappa_2$ and $\kappa_3$ have opposite sign. This property implies that a ghost will always be present in the even-modes sector, as the manifold approaches the FLRW limit. Furthermore, both $\kappa_2$ and $\kappa_3$ vanish in the exact FLRW case. One could wonder
whether these modes, if their mass is finite -- but non-zero, can be integrated out in this same FLRW limit. 
If the masses are heavy, the corresponding modes can be integrated out and the ghost is harmless in general. Otherwise,
the ghost will be physical and the theory -- at least on FLRW backgrounds -- will not be consistent.

In fact, we find
\begin{eqnarray}
\omega_1^2 &\simeq& p^2+ M_{\rm GW}^2 \,,\nonumber\\
\omega_2^2 &\simeq& \!\left(\frac{r^2-1}{24\sigma}\right)\!\!\left[\sqrt{(10p^2+p_T^2)^2 - 8p_L^2p_T^2} - (2p^2 +3p_T^2) \right]\!,\nonumber\\
\omega_3^2 &\simeq& -\omega_2^2+\frac{1-r^2}{12\sigma}\,(2p_L^2+5p_T^2)\,,
\end{eqnarray}
with $\omega_2^2\omega_3^2<0$ in general. Since there is not any mass-gap
(i.e.\ there always exists some value of the momenta for which the
frequency vanishes) for $\omega^2_{2,3}$, we conclude that the ghost is
physical and cannot be integrated out from the Lagrangian. Therefore the
FLRW background is not viable in the dRGT massive gravity theory. Notice
that the first mode corresponds to the massive gravitational
wave. Even though the theory succeeds in removing the BD ghost and giving
the tensor mode a mass, it does not accept a stable FLRW solution. This result agrees with the non-linear analysis of Ref.\cite{D'Amico:2012pi}, where the cubic kinetic terms are shown to be non-vanishing.

We conclude this section by studying the odd modes. For these modes, 
a procedure similar to the one followed for the even modes 
leads to two independent fields. Therefore we confirm the expected presence of 5 dynamical
degrees of freedom for this theory (3 even modes + 2 odd ones). The kinetic terms, and the frequencies for these two independent odd modes read
\begin{equation}
\kappa_{1} = \frac{a^4\,p_L^2\,p_T^4}{2\,p^2}\,,\;\kappa_{2} = \frac{a^4\,p_T^2\,M_{GW}^2}{4\,(1-r^2)}\,\sigma\,,\;\omega^2_{1} = p^2+M^2_{\rm GW}\,,\;\omega^2_{2} = c_{\rm odd}^2\,p^2\,,
\end{equation}
where $c_{\rm odd}^2 = (1-r^2)/(2\sigma)$. Therefore we find a massive tensor mode, and an healthy massless propagating mode (at speed $c_{\rm odd}$), provided that $(1-r)\sigma>0$.

\section{Towards healthy massive cosmologies}
\label{sec:alternatives}

The appearance of the non-linear ghost shown in
subsection~\ref{subsec:nonilnear} originates from the fact that
quadratic kinetic terms exactly vanish: the kinetic terms show up at the
cubic order and can become negative. The disappearance of kinetic terms 
at the quadratic order was shown upon using the background equation of 
motion for the St\"uckelberg fields but without using other background
equations. One can actually show that the off-shell quadratic kinetic
terms have coefficients proportional to 
$J_{\phi}\equiv (3-2X)+\alpha_3(3-X)(1-X)+\alpha_4(1-X)^2$, where 
$X\equiv \alpha/a$, and that the self-accelerating FLRW solution is
characterized by $J_{\phi}=0$ (or $X=X_{\pm}$ with $X_{\pm}$ shown in 
(\ref{eqn:f-sol})).

For this reason, in order to find a stable cosmological background, one
needs to detune the proportionality between the quadratic kinetic terms
and the St\"uckelberg equation of motion characterizing the
self-accelerating background. One way to achieve this would be to relax
the FLRW symmetry by a deformation of the background. This possibility
will be considered in subsection~\ref{sec:anisFLRW}. We shall find that
relatively large deformation by anisotropy in the hidden sector
(fiducial metric) may render the background solution stable. Another
possibility would be to maintain the FLRW symmetry but to change the
St\"uckelberg equation of motion by adding extra dynamical degrees of
freedom to the theory. We shall thus consider the quasi-dilaton
extension in subsection~\ref{subsec:qmg} and the varying mass extension
in subsection~\ref{subsec:vmt}. While self-accelerating FLRW solutions
in the quasi-dilaton theory turn out to be unstable, the varying mass
case allows some stable regimes of parameters. 

Before presenting our results, we note that other extensions, such as the bi-metric theory \cite{bimetric}, where both metrics are dynamical, may also give rise to FLRW type cosmologies \cite{vonStrauss:2011mq, Comelli:2011zm, Akrami:2012vf}, although perturbation analysis in Ref.\cite{Comelli:2012db} indicates that such cosmologies, in the presence of perfect fluids, may develop instabilities.

\subsection{Anisotropic FLRW solution}
\label{sec:anisFLRW}

As argued above, the appearance of the non-linear ghost shown in 
subsection~\ref{subsec:nonilnear} is a consequence of the FLRW
symmetry and the structure of the theory; in order to obtain a stable
solution within the same theory, the FLRW symmetry needs to be relaxed.

An inhomogeneous background solution was obtained in
Ref.~\cite{D'Amico:2011jj}, where the observable universe is
approximately FLRW for a horizon size smaller than the Compton length of
graviton. Similar solutions with inhomogeneities in the St\"uckelberg 
sector, meaning that the physical metric and the fiducial metric do not
have common isometries acting transitively, were found in
\cite{brokenFRW}. Note that those inhomogeneous solutions cannot 
be isotropic everywhere since isotropy at every point implies
homogeneity. Note also that cosmological perturbations can in principle
probe inhomogeneities in the St\"uckelberg sector. For example, generic
spherically-symmetric solutions are isotropic only when they are
observed from the origin.

The goal of this subsection is, following \cite{Gumrukcuoglu:2012aa}, to
introduce an alternative option, where the assumption of isotropy is
dropped but homogeneity is kept. In a
region with relatively large anisotropy, we find an attractor
solution. On the attractor, the physical metric is still isotropic, and
the background geometry is of FLRW type. Hence, the thermal history of
the standard cosmology can be accommodated in this class of
solutions. However, the St\"uckelberg field configuration is
anisotropic, which may lead to effects at the level of the
perturbations, suppressed by smallness of the graviton mass.

\subsubsection{Fixed point solutions}

In this subsection, we review anisotropic FLRW solutions in the dRGT
theory with the de Sitter reference
metric~\cite{Gumrukcuoglu:2012aa}.\footnote{The adoption of the de Sitter reference metric here is due to the flat spatial curvature associated with the Bianchi type--I metric. We remind the reader that the Minkowski reference metric cannot be put into a flat FLRW form (see Sec.\ref{subsec:openFLRW}).}

 The fiducial metric is obtained by
taking (\ref{eqn:generalfid}) and setting $K=0$ and
$H_f\equiv\dot{\alpha}/(n\,\alpha)={\rm constant}$. For the physical 
metric we consider axisymmetric anisotropic extension of a flat FLRW,
i.e.\ the axisymmetric Bianchi type-I metric, given by 
\begin{equation}
 g^{(0)}_{\mu\nu}dx^{\mu}dx^{\nu} =
  -dt^2 + a^2(t)
  [e^{4\sigma(t)}dx^2+e^{-2\sigma(t)}\delta_{ij}dy^idy^j]\,,
\end{equation}
where indices $i,j=2,3$ correspond to the coordinates on the $y{-}z$ plane.

The St\"uckelberg equation gives 
\begin{equation}
 J_\phi^{(x)}\,\left(H+2\,\Sigma -H_f\,e^{-2\,\sigma}\,X\right)
+2\,J_\phi^{(y)}\,\left(H-\Sigma-H_f\,e^{\sigma}\,X\right)=0\,,
\label{anis-stuckfieldeq}
\end{equation}
where
\begin{eqnarray}
J_\phi^{(x)}  &\equiv& (3+3\,\alpha_3+\alpha_4)-2\,(1+2\,\alpha_3+\alpha_4)\,e^\sigma\,X+(\alpha_3+\alpha_4)\,e^{2\sigma}\,X^2\,,\\
J_\phi^{(y)} &\equiv& (3+3\alpha_3+\alpha_4)-(1+2\,\alpha_3+\alpha_4)(e^{-2\sigma}+e^\sigma)X+(\alpha_3+\alpha_4)e^{-\sigma}X^2\,,\nonumber
\end{eqnarray}
$H\equiv \dot{a}/a$, 
 $\Sigma\equiv\dot{\sigma}$ and
$X\equiv \alpha/a$. The metric equations of motion are given by
\begin{eqnarray}
&&3\left(H^2 - \Sigma^2\right)-\Lambda = m_g^2\left[ -(6+4\alpha_3+\alpha_4)+
 (3+3\,\alpha_3+\alpha_4)\,(2\,e^\sigma +e^{-2\sigma})X\right.
\nonumber\\
&&\left.\qquad\qquad\qquad\qquad\quad- (1+2\,\alpha_3+\alpha_4)(e^{2\sigma}+2\,e^{-\sigma})\,X^2 +(\alpha_3+\alpha_4)\,X^3\right],\nonumber\\
&&\dot{\Sigma}+3H\Sigma =
\frac{m_g^2}{3} (e^{-2\,\sigma}-e^\sigma)X\left[ (3+3\,\alpha_3+\alpha_4) -(1+2\,\alpha_3+\alpha_4)(e^\sigma+r)X
\right.
\nonumber\\
&&\left.\qquad\qquad\qquad\qquad\qquad\qquad\qquad\qquad\qquad
+ (\alpha_3+\alpha_4)\,r e^\sigma X^2\right]\,,
\label{eqeins}
\end{eqnarray}
where 
\begin{equation}
r \equiv \frac{n\,a}{\alpha}\equiv \frac{1}{X\,H_f}\,\left(\frac{\dot{X}}{X}+H\right)\,.
\end{equation}
We now look for solutions that are anisotropic
($\Sigma=0,\,\sigma=\sigma_0\ne 0$) and undergo a de Sitter
($\dot{H}=0,\,H=H_0$) expansion, which implies that the remaining
parameters are also constant, i.e.\ $X=X_0,\,r=r_0$. Excluding the fixed
points which are isotropic $\sigma_0=0$ and those which exist only for
special values of parameters (i.e.\ in a measure-zero subspace of the
parameter space), we find a characteristic relation for the anisotropic 
fixed point, $r_0=e^{-2\sigma_0}$, or, 
\begin{equation}
X_0 = \frac{H_0}{H_f}\,e^{2\,\sigma_0}\,.
\label{fixedpoint}
\end{equation}
The remaining two equations allow us to determine $X_0$ and
$\sigma_0$. One can show that this solution is stable against
homogeneous linear perturbations if \cite{Gumrukcuoglu:2012aa}
\begin{equation}
M_\sigma^2 \equiv -\frac{3\,M^2\,\tilde{M}^2\,(9\,H_0^2+\tilde{M}^2)}{\tilde{M}^4+9\,H_0^2(3\,M^2-\tilde{M}^2)} > 0\,,
\label{msigma}
\end{equation}
where \cite{DeFelice:2013awa}
\begin{eqnarray}
M^2 &\equiv& \frac{H_0\,m_g^2}{3\,H_f^3}\,\left[H_0^2 e^{3\,\sigma_0}(1+2\,e^{3\,\sigma_0})(\alpha_3+\alpha_4) - 2\,H_0\,H_f\,(1+e^{3\,\sigma_0}+e^{6\,\sigma_0})\right.\nonumber\\
&&\quad\qquad\left.\times (1+2\,\alpha_3+\alpha_4) + H_f^2(2+e^{3\,\sigma_0})(3+3\,\alpha_3+\alpha_4)\right]\,,\nonumber\\
\tilde{M}^2 &\equiv &-\frac{3\,H_0\,m_g^2}{2\,H_f^3}\,
\left[H_0^2 e^{6\,\sigma_0}(\alpha_3+\alpha_4) -
 2\,H_0\,H_f\,e^{3\,\sigma_0}(1+2\,\alpha_3+\alpha_4)\right.\nonumber\\
&& \qquad\qquad\qquad\left. + H_f^2(3+3\,\alpha_3+\alpha_4)\right]\,.
\label{eqn:Mtilde}
\end{eqnarray}

\subsubsection{Linear perturbations}

The linear perturbation theory around axisymmetric Bianchi type-I
backgrounds in dRGT theory was formulated in \cite{DeFelice:2013awa}. The
formulation can be used to calculate the coefficients of the kinetic terms
for the five gravity degrees of freedom. Similarly to the off-shell kinetic
terms around isotropic FLRW solutions mentioned in the second paragraph
of Sec.~\ref{sec:alternatives}, it turns out that the kinetic term of
one of the five degrees is proportional to $J_\phi^{(x)}$ and other two are
proportional to $J_\phi^{(y)}$. The remaining two correspond to the
standard two polarizations of the tensor graviton and thus they always
have finite and positive kinetic terms.

By studying (\ref{anis-stuckfieldeq}), we see that on the fixed point
(\ref{fixedpoint}), $J_\phi^{(x)} \neq 0$ while $J_\phi^{(y)}=0$. 
Hence, the kinetic terms for two of the expected five gravity degrees of freedom
vanish, signaling for necessity of non-linear analysis.

\subsubsection{Non-linear perturbations}

Due to the broken $SO(3)$ symmetry, we can no longer use the standard 
scalar/vector/tensor decomposition for the perturbations. However, the
axisymmetry of the background allows us to use $SO(2)$ symmetry in the
classification. The following analysis is based on
Ref.~\cite{DeFelice:2013awa}.

{\it Even modes.---}
The even mode perturbations are introduced according to the decomposition (\ref{eq:evenmetric}) and once the non-dynamical degrees are integrated out, there are generically three dynamical degrees of freedom. However, once the background is fixed to be the anisotropic attractor solution, one of these modes has a vanishing kinetic term. On the other hand, we can still analyze the properties of the higher order kinetic terms by considering homogeneous deformations around the attractor solution, characterized by
\begin{equation}
\sigma = \sigma_0 + \epsilon \,\sigma_1 +{\cal O}(\epsilon^2)\,,\qquad
\Sigma = \epsilon\,\Sigma_1 +{\cal O}(\epsilon^2) = \epsilon\,\dot{\sigma_1}+{\cal O}(\epsilon^2)\,,
\end{equation}
where the background-physical-metric coefficients are given by $g_{tt}=-N^2$, $g_{xx}=a^2e^{4\sigma}$, $g_{yy}=g_{zz}=a^2 e^{-2\sigma}$. Furthermore, we have defined $\Sigma$ as $\Sigma \equiv \dot{\sigma}$,
 and expanded both $\sigma$ and $\Sigma$ on the attractor solution.
After diagonalization, the kinetic terms become
\begin{eqnarray}
\kappa_1 &\simeq& \left[ \frac{8\,p^4}{p_T^4} - \frac{8\,\tilde{M}^4}{\tilde{M}^4+9\,H_0^2(3\,M^2-\tilde{M}^2)}\right]^{-1}\,,\\
\kappa_2 &\simeq& \frac{2\,a_0^4\,e^{8\,\sigma_0}\,\tilde{M}^2\,p_L^2\,\left[9\,H_0^2\,p^4\,(\tilde{M}^2-3\,M^2)+\tilde{M}^4\,p_L^2\,(-2\,p^2+p_L^2)\right]}{\tilde{M}^2\,p_L^2(\tilde{M}^2-3\,p^2)^2-9 H_0^2(\tilde{M}^2-3\,M^2)\,\left[6\,p^4+\tilde{M}^2(-4\,p^2+p_L^2)\right]}\,,\nonumber\\
\label{eq:kinvan}\kappa_3 &\simeq& -\frac{3\,\tilde{M}^2\,e^{2\,\sigma_0}\,a_0^4\,p_T^2\,\left[3\,M^2\,(9\,H_0^2+\tilde{M}^2)\sigma_1 + 2\,H_0\,(9\,M^2-2\,\tilde{M}^2)\Sigma_1\right]}{(1-e^{6\,\sigma_0})\,\left[\tilde{M}^4-27\,H_0^2\,(3\,M^2-\tilde{M}^2)\right]}\,,\nonumber
\label{even-kappas}
\end{eqnarray}
where $p_L$ and $p_T$ are the components of the physical momentum vector along the $\hat{x}$ direction and on the $y{-}z$ plane, 
respectively, while $p^2\equiv p_L^2+p_T^2$. Furthermore we have introduced the mass scale $\tilde M^2$ as in Eq.~(\ref{eqn:Mtilde}). Generically, the absence of ghosts imposes momentum dependent conditions. However, one can ensure stability at all scales by adopting the sufficient condition
\begin{equation}
\tilde{M}^2 < 0 \,, \qquad
M^2 < \frac{\tilde{M}^2(9 H_0^2-\tilde{M}^2)}{27\,H_0^2} < 0\,,
\label{noghost1}
\end{equation}
under which, both $\kappa_1$ and $\kappa_2$ can be made positive. For a
parameter set which satisfies (\ref{noghost1}), the no-ghost condition
for the third mode (with order $\epsilon$ kinetic term) becomes
\begin{equation}
\sigma_0\left[-3\vert M^2\vert\left(9\,H_0^2-\vert \tilde{M}^2\vert\right)\sigma_1 +2\,H_0\,\left(2\vert \tilde{M}^2\vert-9\vert M^2\vert\right)\Sigma_1\right]<0\,,
\label{noghost3}
\end{equation}
which depends linearly on the homogeneous deformations $\sigma_1$ and $\Sigma_1$ around the fixed point. Thus, regardless of the value of $M^2$ and $\tilde{M}^2$, there could always be a region where $\sigma_1$ and $\Sigma_1$ conspire to render the third mode a ghost. On the other hand, if the initial conditions are such that the system is close to the attractor, it is possible to connect the evolution of $\Sigma_1$ algebraically to that of $\sigma_1$ and obtain a regime where one can avoid the instability. By considering the equation of motion for $\sigma_1$,
\begin{equation}
\dot{\Sigma}_1 +3\,H_0 \,\Sigma_1 + M_\sigma^2\,\sigma_1 = 0\,,
\label{eq:sigma1evol}
\end{equation}
we first note that the condition (\ref{msigma}) for the stability of the
fixed point against homogeneous perturbations, combined with the
conditions (\ref{noghost1}), yields 
\begin{equation}
9 H_0^2 - \vert\tilde{M}^2\vert >0\,.
\label{fixedpointstability}
\end{equation}
To satisfy the condition (\ref{noghost3}), we suppose that the system is 
in the attractor regime, so that $\Sigma_1\propto \sigma_1$ and that
$\sigma_1$ does not change sign during the course of evolution. This
scenario can be attained if the friction term in (\ref{eq:sigma1evol})
dominates over the mass term, i.e.\ 
\begin{equation}
9\,H_0^2> 4\,M_\sigma^2\,.
\label{noghost2}
\end{equation}
Then, solving Eq.~(\ref{eq:sigma1evol}) and evaluating the solution at late times, we find the relation
\begin{equation}
\Sigma_1  \simeq \left(-\frac{3}{2}H_0 + \sqrt{\frac{9}{4}\,H_0^2-M_\sigma^2}\right)\sigma_1\,.
\label{lateattractor}
\end{equation}
Thus, in this regime, the condition (\ref{noghost3}) can in principle be satisfied by choosing the appropriate sign for $\sigma_1$.

{\it Odd modes.---} 
The odd mode perturbations are introduced according to the decomposition (\ref{eq:oddmetric}) and once the non-dynamical degree is integrated out (another mode can be gauged away), there are generically two dynamical degrees of freedom. On the anisotropic attractor solution, the kinetic term of one of these modes vanishes, and as we did for the even modes, we consider homogeneous deviations from the fixed point to determine the conditions for non-linear stability.

After diagonalization, the kinetic terms become
\begin{eqnarray}
\kappa_{1} &\simeq& \frac{a_0^4\,e^{-4\,\sigma_0}\,p_L^2\,p_T^4}{2\,p^2}\,,\\
\kappa_{2} &\simeq& -\frac{3\,\tilde{M}^2\,e^{2\,\sigma_0}\,a_0^4\,p_T^2\,\left[3\,M^2\,(9\,H_0^2+\tilde{M}^2)\sigma_1 + 2\,H_0\,(9\,M^2-2\,\tilde{M}^2)\Sigma_1\right]}{4\,(1-e^{6\,\sigma_0})\,\left[\tilde{M}^4-27\,H_0^2\,(3\,M^2-\tilde{M}^2)\right]}\,.\nonumber
\end{eqnarray}
The first kinetic term is always positive, whereas the second mode acquires a kinetic term proportional to the deviation from the fixed point. In fact, up to a numerical factor, $\kappa_2$ above is the same as the kinetic term of the third mode in Eq.~(\ref{even-kappas}), so if the conditions discussed in the even sector are satisfied, the odd sector will also be stable.

\subsection{Extended theory I: quasi-dilaton}
\label{subsec:qmg}

The quasi-dilaton theory is obtained by introducing a scalar field
$\sigma$ associated with a dilation-like global symmetry to the dRGT action, ($\sigma$ has
different meaning than the previous subsection) 
\begin{equation}
\sigma \to \sigma-\alpha M_{\rm Pl}\,,
\qquad
\phi^a \to e^{\alpha}\,\phi^a\,.
\end{equation}
The action compatible with this symmetry is given in Einstein frame as
\cite{D'Amico:2012zv} 
\footnote{There is an additional term allowed by the symmetry, $\int d^4x\,\sqrt{-f}\,e^{4\,\sigma/M_p}$, which does not change the conclusions of the present discussion (See \cite{Gumrukcuoglu:2013nza, D'Amico:2013kya} for details).}.
\begin{eqnarray}
S &=& \frac{M_{\rm Pl}^2}{2}\int\,d^4x\,\sqrt{-g}\,\left[R - 2\,\Lambda+ 2\,m_g^2 \left({\cal L}_2(\bar{{\cal K}}) +\alpha_3\,{\cal L}_3(\bar{{\cal K}}) +\alpha_4\,{\cal L}_4(\bar{{\cal K}})\right) \right.\nonumber\\
&&\left.\qquad\qquad\qquad\qquad\qquad\qquad\qquad
-\frac{\omega}{M_{\rm Pl}^2}\,\partial_\mu\sigma\,\partial^\nu\sigma + {\cal L}_{\rm matter}\right]\,,
\end{eqnarray}
where ${\cal L}_2$, ${\cal L}_3$ and ${\cal L}_4$ are given in
Eq.~(\ref{lag234}), but the building block tensor (\ref{Kdef}) is
replaced with 
\begin{equation}
{\cal K}^\mu_\nu \to \bar{{\cal K}}^\mu_\nu \equiv \delta^\mu_\nu-e^{\sigma/M_{\rm Pl}} \left(\sqrt{g^{-1}f}\right)^\mu_\nu\,.
\label{barKdef}
\end{equation}

\subsubsection{Self-accelerating solutions}

We adopt the Minkowski reference metric and the flat FLRW ansatz for the
physical metric as 
\begin{equation}
f_{\mu\nu} = -n^2(t)\delta^0_\mu \delta^0_\nu
 +\delta_{ij}\delta^i_{\mu}\delta^j_{\nu}\,,\quad
g_{\mu\nu} dx^\mu\,dx^\nu = -dt^2+a^2(t)\,\delta_{ij} dx^i\,dx^j\,.
\label{flatfrw}
\end{equation}
The equations of motion for the St\"uckelberg fields yield
\begin{equation}
(1-X)\,X\,
\left[
3+3\,\alpha_3+\alpha_4- (3\,\alpha_3+2\,\alpha_4)  \,X + \alpha_4\,X\,^2
\right] =
\frac{{\rm constant}}{a^4},
\label{eqstuck}
\end{equation}
where $X\equiv e^{\sigma/Mp}/a$, leading to four attractors: $X=0$,
$X=1$ and $X=X_{\pm}$ with 
\begin{equation}
X_\pm =
\frac{3\,\alpha_3+2\,\alpha_4 \pm \sqrt{9\,\alpha_3^2
-12\,\alpha_4}}{2\,\alpha_4}\,. 
\label{xpm}
\end{equation}
Among them, $X=0$ and $X=1$ leads to either strong coupling or
instability~\cite{D'Amico:2012zv}. We thus consider $X=X_{\pm}$ as
backgrounds. Along these branches of solutions, the (modified) Friedmann
equation becomes 
\begin{equation}
\left(3 - \frac{\omega}{2}\right)H^2 = \Lambda +\Lambda_\pm\,,
\label{qmg-friedmann}
\end{equation}
where the graviton mass manifests as the effective cosmological constant
\begin{equation}
\Lambda_\pm =-\frac{m_g^2}{2\,\alpha_4^3}\,\left[9\,(3\,\alpha_3^4-6\,\alpha_3^2\,\alpha_4 +2\,\alpha_4^2) \pm \alpha_3\,(9\,\alpha_3^2-12\,\alpha_4)^{3/2}\right]
\,.
\end{equation}
 From (\ref{qmg-friedmann}), we immediately see that a sensible
cosmology requires $\omega<6$. Finally, the equation of motion for the
quasi-dilaton field gives 
\begin{equation}
r \equiv n \,a= 1+ \frac{\omega\,H^2}{m_g^2\,X^2\left[\alpha_3\,(X - 1) -2\right]}\,.
\end{equation}

\subsubsection{Perturbations}

We now introduce perturbations as \cite{Gumrukcuoglu:2013nza}
\begin{eqnarray}
\sigma & = & M_{\rm Pl} \left[\log(a\,X)+\delta\sigma\right]\,, \quad
\delta g_{00} = -2\,\Phi\,,\quad
\delta g_{0i} = a (B_i^T+\partial_iB)\,,\nonumber\\
\delta g_{ij} &=& a^2\left[2\,\Psi\,\delta_{ij} + \left(\partial_i\partial_j - \frac{\delta_{ij}}{3}\,\partial_l\partial^l\right)E+\frac{1}{2} (\partial_i E^T_j +\partial_j E^T_i) +h_{ij}^{TT}\right]\,,
\label{metricdecomp}
\end{eqnarray}
while we fix the unitary gauge $\delta\phi^a=0$, where $B_i^T$ and
$E^T_i$ are transverse and $h_{ij}^{TT}$ is transverse and traceless. With
respect to the dRGT theory, we have an additional scalar field, so in
total, we  expect $2$ tensor, $2$ vector and $2$ scalar degrees, once
the non-dynamical modes are integrated out.

{\it Tensor perturbations.---}
The quadratic action for the tensor modes reduces to
\begin{equation}
S_{\rm T}= \frac{M_{\rm Pl}^2}{8}\,\int d^3k\,a^3\,dt\,\left[|\dot{h}^{TT}_{ij}|^2-\left(\frac{k^2}{a^2}+M_{GW}^2\right)|h^{TT}_{ij}|^2\right]\,,
\end{equation}
where
\begin{equation}
M_{GW}^2 \equiv  \frac{m^2_g\,(r-1)\,X^3}{X-1} +H^2\,\omega\,\left(\frac{r}{r-1}+\frac{2}{X-1}\right)\,.
\label{mgwqmgdef}
\end{equation}
Generically, $M_{GW}\sim {\cal O}(H)$ so even if the tensor modes are
tachyonic, the time scale of their instability is of the order of the
age of the universe~\cite{Gumrukcuoglu:2013nza}. 

{\it Vector perturbations.---}
For the vector modes, the quadratic action is
\begin{equation}
S_{\rm V}= \frac{M_{\rm Pl}^2}{16}\,\int d^3k\,a^3\,dt\,k^2\left[\frac{|\dot{E}^T_{i}|^2}{\left(1+ \frac{k^2(r^2-1)}{2a^2\,H^2\,\omega}\right)} -M_{GW}^2|E^T_{i}|^2\right]\,.
\label{qmg-vec-act1}
\end{equation}
We see from (\ref{qmg-vec-act1}) that if $(r^2-1)/\omega <0$, there is a
critical momentum above which the vector modes have ghost
instability. Therefore, the UV cutoff scale of the effective theory
$\Lambda_{UV}$ should be lower than this critical (physical) momentum to
ensure the stability of the system: 
$\Lambda_{UV}^2\,(1-r^2)/(H^2\,\omega)<2$. In addition, the frequency in the canonical normalization yields a further condition on avoiding tachyonic instability, which arises if $M_{GW}^2>0$ and $(r^2-1)/\omega >0$. The growth rate of this instability can be made lower than or
at most of the cosmological scale for all physical momenta below the UV
cut-off $\Lambda_{UV}$, provided that \cite{Gumrukcuoglu:2013nza}
\begin{equation}
\Lambda_{UV}^2 \lesssim \frac{2\,H^2\,\omega}{r^2-1}
\,.
\end{equation}

{\it Scalar perturbations.---}
After integrating out $\delta g_{0\,\mu}$ as well as the would-be BD degree, the scalar sector contains two coupled modes. The 
kinetic part of the quadratic action is formally written as
\begin{equation}
S_{\rm S} \ni \int \frac{d^3k}{2}\,a^3\,dt\,\left[K_{11} \vert\dot{Y_1}\vert^2+ K_{22} \vert\dot{Y_2}\vert^2+ K_{12} \left(\dot{Y_1}\, \dot{Y_2}^\star + \dot{Y_2} \,\dot{Y_1}^\star \right) \right]\,,
\end{equation}
where $Y_1$ and $Y_2$ are particular linear combinations of $\Psi$ and
$\delta\sigma$. For our purposes, it is enough to study the determinant
of the kinetic matrix, given by 
\begin{equation}
 \det K \equiv 
K_{11}K_{22}-K_{12}^2 = \frac{3\,k^6\,\omega^2\,a^4\,H^4}{\left[\omega\,a^2\,H^2- \frac{4\,k^2}{(6-\omega)}\right]\,(r-1)^2}\,,
\label{detK}
\end{equation}
The absence of ghost degrees in the scalar sector requires $\det K>0$ as
a necessary condition. We first note that the determinant is always
negative if $\omega<0$. Along with the condition obtained from
(\ref{qmg-friedmann}), we thus obtain $0<\omega <6$ as a necessary 
condition. 

Furthermore, demanding that $\det K>0$ for all physical momenta below
the UV cutoff of the theory, we obtain
\begin{equation}
\frac{\Lambda_{UV}}{H} < \frac{\sqrt{\omega (6-\omega)}}{2}
< \frac{3}{2}\,,
\label{noghostsca}
\end{equation}
where we have used the condition $0<\omega<6$ to obtain the last
inequality. Unfortunately, (\ref{noghostsca}) is not acceptable since it
would imply that the UV cutoff scale would be lower than the
cosmological scale and that the theory would not be applicable to
cosmology. Therefore, we conclude that for physical wavelengths shorter
than cosmological scales, $\det K<0$ and one of the two degrees of
freedom is a ghost~\cite{Gumrukcuoglu:2013nza}.

It can also be checked (see \cite{Gumrukcuoglu:2013nza} for details) that
energies of the ghost mode are not parametrically higher than 
$H\sim m_g$. This signals the presence of ghost instabilities in the
regime of validity of the effective field theory.

\subsection{Extended theory II: varying mass}
\label{subsec:vmt}

A further way of extending the dRGT theory is to allow the parameters of
the theory to vary with a scalar field $\sigma$. The action in this case
is 
\cite{Huang:2012pe}, 
\begin{eqnarray}
S&=& \int\,d^4x\,\sqrt{-g}\,\left\{M_{\rm Pl}^2\Big[\frac{R}{2}- \Lambda+ m_g^2(\sigma) \left[{\cal L}_2 +\alpha_3(\sigma)\,{\cal L}_3 +\alpha_4(\sigma)\,{\cal L}_4\right]\Big]\right.\nonumber\\
&&\left.\qquad\qquad\qquad\qquad\qquad\qquad\qquad
-\frac{1}{2}\,\partial_\mu\sigma\,\partial^\nu\sigma-V(\sigma) + {\cal L}_m\right\}\,,
\end{eqnarray}
where ${\cal L}_2$, ${\cal L}_3$ and ${\cal L}_4$ are given by Eq.~(\ref{lag234}).

\subsubsection{Background}

As in the previous subsection, we adopt the Minkowski reference metric
and the flat FLRW ansatz for the physical metric (\ref{flatfrw}). 
The equations of motion for the St\"uckelberg fields yield
\begin{equation}
\frac{m_g^2
(X-1)}{X^3}\,
\left[
3-3\,(X-1)\,\alpha_3+(X-1)^2\,\alpha_4\right]
= {\rm constant},
\label{vm-eqstuck}
\end{equation}
where $X \equiv 1/a$ and $r \equiv a\,n$. Due to the assumptions of flat
space and Minkowski reference metric, if $m_g$ and $\alpha_{3,4}$ are
time-independent, the solution $X={\rm constant}$ does not allow any 
non-trivial cosmologies (see the second paragraph of
subsection~\ref{subsec:openFLRW}). 

By defining 
\begin{eqnarray}
\rho_m &\equiv& M_{\rm Pl}^2 m_g^2\,(X-1)\,\left[6+4\,\alpha_3+\alpha_4-X\,(3+5\,\alpha_3+2\,\alpha_4)+X^2\,(\alpha_3+\alpha_4)\right]\,,
\nonumber\\
p_m &\equiv& M_{\rm Pl}^2 m_g^2\,\left[6+4\,\alpha_3+\alpha_4-(2+r)\,X\,(3+3\,\alpha_3+\alpha_4)\right.\nonumber\\
&&\left.\qquad\qquad\qquad\qquad
+(1+2\,r)\,X^2\,(1+2\,\alpha_3+\alpha_4)-r\,X^3\,(\alpha_3+\alpha_4)\right]\,,
\nonumber\\
Q&\equiv& 
M_{\rm Pl}^2\,m_g^2\,\dot{\sigma}\,(X-1)^2\Bigg\{
\alpha_3'\,(4-X-3\,r\,X)+\alpha_4'(X-1)\,(r\,X-1)
\nonumber\\
&&+\frac{2\,m_g'}{m_g}\,
\Bigg[3-(X-1)\alpha_3+ \frac{r\,X-1}{X-1}\left[3-3\,(X-1)\,\alpha_3 +(X-1)^2\alpha_4\right]\Bigg]\Bigg\}\,,
\nonumber\\
\rho_\sigma &\equiv&  \frac{\dot{\sigma}^2}{2}+V \,,\quad
p_\sigma \equiv \frac{\dot{\sigma}^2}{2}-V \,,
\end{eqnarray}
we can write the set of background equations of motion in the following form
\begin{eqnarray}
&&3\,H^2 = \Lambda+ \frac{1}{M_{\rm Pl}^2}\,\left(\rho_\sigma + \rho_m \right)\,, \quad
\dot{H} = -\frac{1}{2\,M_{\rm Pl}^2}\,\left[ (\rho_\sigma+p_\sigma) +(\rho_m+p_m)\right]\,, \nonumber\\
&& \dot{\rho}_m+3\,H\,(\rho_m+p_m)=-Q\,,\quad
 \dot{\rho}_\sigma+3\,H\,(\rho_\sigma+p_\sigma)=Q\,.
\end{eqnarray}
where prime denotes differentiation with respect to $\sigma$.

Although dynamical analysis for these equations have been studied in the
literature \cite{ExtCos}, there is not yet a simple
self-accelerating solution in the varying parameter massive gravity. In
the following, we do not assume any specific evolution and keep the
functions $m_g(\sigma)$, $\alpha_3(\sigma)$ and $\alpha_4(\sigma)$
generic.

\subsubsection{Perturbations}

We now introduce perturbations, following \cite{Gumrukcuoglu:2013nza}. The
metric is decomposed as in Eq.~(\ref{metricdecomp}) and we adopt the
unitary gauge as $\delta\phi^a=0$, while the scalar field is perturbed
as
\begin{equation}
\sigma = \langle \sigma \rangle +M_{\rm Pl}\,\delta\sigma\,.
\end{equation}

{\it Tensor perturbations.---}
The tensor action reduces to
\begin{equation}
S_{\rm T}= \frac{M_{\rm Pl}^2}{8}\,\int d^3k\,a^3\,dt\,\left[|\dot{h}^{TT}_{ij}|^2-\left(\frac{k^2}{a^2}+M_{GW}^2\right)|h^{TT}_{ij}|^2\right]\,,
\end{equation}
where
\begin{equation}
M_{GW}^2=\frac{(r-1)\,X^2}{(X-1)^2}\,\left[m_g^2\,(X-1) - \frac{\rho_m}{M_{\rm Pl}^2}\right]
-\left(\frac{1}{r-1}+\frac{2\,X}{X-1}\right)
\,\frac{\rho_m+p_m}{M_{\rm Pl}^2}\,.
\label{vm-mgwdef}
\end{equation}
The stability of long wavelength tensor modes is ensured by $M_{GW}^2>0$.

{\it Vector perturbations.---}
For the vector modes, the action is
\begin{equation}
S_{\rm V}= \frac{M_{\rm Pl}^2}{16}\,\int d^3k\,a^3\,dt\,k^2\left[\frac{|\dot{E}^T_{i}|^2}{\left(1-\frac{k^2\,(r^2-1)\,M_{\rm Pl}^2}{2a^2\,(\rho_m+p_m)}\right)}
-M_{GW}^2|E^T_{i}|^2\right]\,.
\label{vm-vec-act1}
\end{equation}
By requiring that the kinetic term is positive for all physical momenta below the cut-off scale of the theory $\Lambda_{UV}$, we obtain the stability condition for the vector modes as
\begin{equation}
\frac{\Lambda_{UV}^2\,(1-r^2)}{H^2R} < 2,\quad
 R \equiv -\frac{\rho_m+p_m}{M_{\rm Pl}^2H^2} \ .
\label{noghostvec-vm}
\end{equation}
Under the above condition, we can further analyze the stability of the vector sector, by introducing a time reparametrization which renders the modes $E^T_i$ canonical, then requiring that their frequency is an increasing function. This procedure yields sufficient (but not necessary) condition for stability
\begin{equation}
 \left[1 + \frac{1}{8NH}\frac{d}{dt}
 \ln\left(\frac{RM_{GW}^2}{r^2-1}\right)\right]\,\frac{\Lambda_{UV}^2(1-r^2)}{H^2R} < \frac{3}{2} + \frac{1}{4NH}\frac{d\,\ln \left(M_{GW}^2\right)}{dt}\,.
\label{stabilityvm}
\end{equation}
%

{\it Scalar perturbations.---}
As in the quasi-dilaton theory, we integrate out the nondynamical
degrees and are left with two coupled modes in the scalar sector. The
kinetic part of the action is formally 
\begin{equation}
S_{\rm S} \ni \int \frac{d^3k}{2}\,a^3\,dt\,\left[K_{11} \vert\dot{Y_1}\vert^2+ K_{22} \vert\dot{Y_2}\vert^2+ K_{12} \left(\dot{Y_1}\, \dot{Y_2}^\star + \dot{Y_2} \,\dot{Y_1}^\star \right) \right]\,,
\end{equation}
where $Y_1$ and $Y_2$ are linear combinations of $\Psi$ and
$\delta\sigma$. For our purposes, it is enough to study 
$\det K=K_{11}K_{22}-K_{12}^2$, whose explicit form is
\begin{equation}
\det K = 
\frac{3\,M_{\rm Pl}^2\,a^2\,k^6\,(\rho_m+p_m)^2\,(\rho_\sigma+p_\sigma-6\,M_{\rm Pl}^2\,H^2)}{(r-1)^2
\left[4\,M_{\rm Pl}^4\,H^2\,\frac{k^2}{a^2}-(\rho_m+p_m)(\rho_\sigma + p_\sigma -6\,M_{\rm Pl}^2\,H^2) \right]}\,.
\label{vm-detK}
\end{equation}
By requiring that the determinant is positive, we see that in order to
avoid a ghost degree of freedom, the momenta in the range 
$0\leq k/a\leq \Lambda_{UV}$ should all satisfy
\begin{equation}
\left(\frac{\rho_\sigma+p_\sigma}{4\,M_{\rm
 Pl}^2\,H^2}-\frac{3}{2}\right)^{-1}\frac{k^2}{a^2} >
\frac{\rho_m+p_m}{M_{\rm Pl}^2}\,.
\label{eq:vm-noghostsca2}
\end{equation}
Explicit diagonalization of the system shows that this condition is
actually a sufficient condition to avoid ghost instabilities in the
scalar sector~\cite{Gumrukcuoglu:2013nza}.

For a background solution which can effectively describe the late time
acceleration, we can assume a de Sitter like expansion, i.e.\ 
$|{\dot H}| \ll H^2$. With these considerations, the stability
requirement for the scalar sector becomes even simpler,  
\begin{equation}
 R + \frac{4}{R-6}\frac{k^2}{H^2a^2} > 0, 
\label{noghostsca-vm-k}
\end{equation}
where $R$ is defined in (\ref{noghostvec-vm}). If indeed all the
physical momenta below the cutoff scale $\Lambda_{UV}$ satisfy
(\ref{noghostsca-vm-k}) and if we suppose $\Lambda_{UV}/H>3/2$ so that
the theory is applicable to cosmological scales, then the no-ghost
condition for scalar perturbations in the regime $|{\dot H}| \ll H^2$
becomes simply 
\begin{equation}
 R > 6. 
\label{noghostsca-vm}
\end{equation}

\section{Summary and Discussion}

The extension of GR by a mass term has been studied for several
decades. Nonetheless, a self-consistent non-linear massive gravity
theory with five propagating degrees of freedom, dubbed the dRGT theory, has been proposed only recently. 

In the present article, we reviewed several cosmological solutions in
the context of the dRGT theory \cite{deRham:2010ik,deRham:2010kj}. We have firstly described open FLRW solutions
with a Minkowski reference metric. By considering a general FLRW-form
fiducial metric, the branch of open FLRW solutions was generalized to
FLRW solutions with general spatial curvature. However, for all of these
FLRW-type cosmological solutions, the kinetic terms of three among five 
gravity degrees of freedom vanish at the level of the quadratic
action. This phenomenon is a consequence of the symmetry of the
FLRW background. 
On analyzing the behavior of the non-linear
perturbations by considering a consistent truncation, it was then shown that there is always at least one ghost
(among the five degrees of freedom) in the gravity sector.

We have then discussed two approaches towards healthy cosmologies in massive
gravity. One proposal is to introduce relatively large anisotropy in the 
configuration of St\"uckelberg fields, which form the hidden sector of the theory. 
We considered the fixed point solution named as ``anisotropic FLRW'', a
solution with the FLRW symmetry in the visible sector (physical metric)
but with anisotropy in the hidden sector. 
Performing a non-linear analysis around the anisotropic fixed point yields that 
anisotropic FLRW solutions can be ghost-free for a range of parameters and
initial conditions. The second proposal discussed here consists of introducing an extra degree of
freedom coupled to the hidden sector. As examples for this possibility, we have considered the quasi-dilaton
theory and the varying mass model. 
For the quasi-dilaton theory, the self-accelerating background turns out to be unstable.
On the other hand, in the varying mass case, there is a regime
of parameters in which a stable cosmological evolution is possible,
although viable self-accelerating solutions yet remain to be found.

Besides the stability investigation, the study of observational signals
from graviton mass, although not included in this review, is also
important. For example, in \cite{Gumrukcuoglu:2012wt}, it was found
that graviton mass may leave a prominent feature with a sharp peak in
the stochastic gravitational wave spectrum. The position and height of
the peak may tell us information about the graviton mass today
and the duration of the inflationary period.  

Last but not least, as a developing field, massive gravity still leaves
many intriguing unsolved questions. One of the most interesting 
questions is the construction of a possible UV completion of massive 
gravity. One of the potential directions to this end would be to seek a
mechanism that realizes the specific structure of the graviton mass term
as a consequence of a spontaneous symmetry breaking. Another important question is the fate of super-luminal
mode~\cite{superluminal} in the gravity sector. It is generically
expected that in the massless limit, observable effects of the
super-luminal mode should disappear and GR should be recovered, provided 
that the mode is excited by a fixed amount of matter source. Thus, it
should be possible to obtain an observational upper bound on the
graviton mass although it is probably not stronger than $m_g<O(H_0)$.

\section*{Acknowledgments}
A.E.G, C.L. and S.M. thank K.~Hinterbichler, S.~Kuroyanagi, N.~Tanahashi
and M.~Trodden for fruitful
collaborations~\cite{Gumrukcuoglu:2013nza,Gumrukcuoglu:2012wt}. The work of
A.E.G, C.L. and S.M. was supported by WPI Initiative, MEXT,
Japan. S.M. also acknowledges the support by Grant-in-Aid for Scientific
Research 24540256 and 21111006.

\end{document}